\documentstyle[preprint,tighten,aps,psfig]{revtex}
\begin{document}
\title{Parity doublets and chiral symmetry restoration
in baryon spectrum}
\author{ L. Ya. Glozman}
\address{ 
 High Energy Accelerator Research Organization (KEK),
Tanashi Branch, Tanashi, Tokyo 188-8501, Japan
\footnote{e-mail: lyg@cleopatra.kfunigraz.ac.at; 
present address: Institute for Theoretical
Physics, University of Graz, Universit\"atsplatz 5, A-8010
Graz, Austria}}
\maketitle

\begin{abstract} 
It is argued that an appearance of the near parity
doublets in the upper part of the light baryon spectrum
is an evidence for the chiral symmetry restoration in
the regime where a typical momentum of quarks is around
the chiral symmetry restoration scale.
At high enough baryon excitation energy the nontrivial gap
solution, which signals the chiral symmetry breaking
regime, disappears and the chiral symmetry should be
restored.
Thus one observes a
phase transition in the upper part of the light baryon
spectrum. The average kinetic energy of the constituent
quarks in this region is just around the critical
one $3T_c$. 
\end{abstract}

\bigskip
\bigskip

The still poorly mapped upper part of the light baryon spectrum
\cite{Caso} exhibits  remarkable parity doublet patterns. To these belong
$N(2220), \frac{9}{2}^+ - N(2250), \frac{9}{2}^-$, 
$N(1990), \frac{7}{2}^+ - N(2190), \frac{7}{2}^-$,
$N(2000), \frac{5}{2}^+ - N(2200), \frac{5}{2}^-$,
$N(1900), \frac{3}{2}^+ - N(2080), \frac{3}{2}^-$,
$\Delta(2300), \frac{9}{2}^+ - \Delta(2400), \frac{9}{2}^-$,
$\Delta(1950), \frac{7}{2}^+ - \Delta(2200), \frac{7}{2}^-$,
$\Delta(1905), \frac{5}{2}^+ - \Delta(1930), \frac{5}{2}^-$,
$\Delta(1920), \frac{3}{2}^+ - \Delta(1940), \frac{3}{2}^-$,
$\Delta(1910), \frac{1}{2}^+ - \Delta(1900), \frac{1}{2}^-$.
 The splittings within the parity partners are typically
within the 5\% of the baryon mass. This value should be given
a large uncertainty range because of the experimental uncertainties
for the baryon masses of the order 100 MeV. Only a couple of states
in this part of the spectrum do not have their parity partners so
far observed. The low energy part of the spectrum, on the
other hand, does not show this property of the parity doubling.
The increasing amount of the near parity doublets in the high
energy sector was a motivation to speculate that the baryon
spectrum exhibits a smooth transition from the Nambu-Goldstone
mode of chiral symmetry in the low-energy part to the Wigner-Weyl
mode in the upper part \cite{GR}. In this note I shall give a
further impetus to this idea and show that this phenomenological
observation has in fact a simple and transparent microscopical
foundation.\\

The almost perfect $SU(2)_L \times SU(2)_R$ global chiral
symmetry of the QCD Lagrangian in the $u,d$ sector is
equivalent to the independent vector and axial rotations
in the isospin space. The axial transformation mixes
states with different spatial parities. Hence, if this
symmetry of the QCD Lagrangian were intact in the vacuum,
one would observe parity degeneracy of all hadron states
with otherwise the same quantum numbers. This is however
not so and it was a reason for  suggestion in the early days of QCD
 that the chiral symmetry of the QCD Lagrangian is broken down
to the vectorial subgroup $SU(2)_V$ by the QCD vacuum, which 
reflects a conservation of the vector current (baryon number). 
That this is so is directly
evidenced by the nonzero value of the quark condensate

\bigskip
\bigskip

\begin{equation}
<\bar \psi \psi> \simeq -(240 - 250 MeV)^3 
\label{condensate},
\end{equation}

\noindent
which represents the order parameter associated with
the chiral symmetry breaking. The nonzero value of the
quark condensate directly shows that the vacuum state
is not chiral-invariant.\\

Physically the nonzero value of the quark condensate implies
that the energy of the state which contains an admixture
of ``particle-hole'' excitations (real vacuum of QCD) is below
the energy of the vacuum for a free Dirac field, in which
case all the negative-energy levels are filled in and all the
positive-energy levels are free. This can happen only due
to some nonperturbative gluonic interactions between quarks
which pairs the left quarks and the right antiquarks (and vice versa)
in the vacuum.
In perturbation theory to any order the structure of the
trivial Dirac vacuum persists. Such a situation is typical in
many-fermion systems (compare, e.g., with the theory of
superconductivity) and implies that there must appear
quasiparticles with dynamical masses. While there are
indications that the instanton fluctuations of the
gluonic field could be important for the chiral
symmetry breaking and thus for the creation of quasiparticles
\cite{DIAKONOV}, the issue on the true dynamical
mechanism of the chiral symmetry breaking in QCD is still
unclear. For example, the nonperturbative resummation of
gluonic exchanges by solving the Schwinger-Dyson equation
\cite{Roberts} is known to also lead to chiral symmetry
breaking if the strong coupling constant is big enough
(however, in the latter case the $U(1)_A$ problem persists),
or it can be generated by monopole condensation, which is
a fashionable scenario for the string-like confining force
in QCD.\\

Formally  the quark condensate represents the closed loop
in momentum space, with the fermion line beginning and ending at
the same space-time point:

\begin{equation}
<\bar \psi \psi> = -i ~tr S_F(0) 
\label{definition},
\end{equation}

\noindent
where $S_F(x-y)$ is Dirac Green function:

\begin{equation}
S_F(x-y) = -i <T[\psi(x) \bar \psi(y)]> = 
\int \frac{d^4 p}{(2\pi)^4} e^{ip(x-y)} \frac{\not p + m}{p^2-m^2+i\epsilon} 
\label{green}.
\end{equation}

\noindent The trace over slash term gives identically zero. Hence
the nonzero value of the quark condensate implies that the
massless quark field ($m=0$) acquires a non-zero dynamical
mass, $M(p)$, which should be in general a momentum-dependent
quantity

\begin{equation}
<\bar \psi \psi> = - 4N_c i  
\int \frac{d^4 p}{(2\pi)^4}  \frac{ M(p)}{p^2-M^2(p)+i\epsilon}
\label{condensate2}.
\end{equation}

\noindent
The dynamical mass $M(p)$ also represents the order parameter.
This  mass should vanish at high momentum, in which
case the quarks are not influenced by the QCD medium and
perturbative QCD is applicable. But at small momenta, below
the chiral symmetry breaking scale, $\Lambda_\chi$, the QCD
nonperturbative phenomena become crucial and give rise to
the nonzero dynamical mass as well as to the condensate. This
dynamical mass  at small momenta can be evidently 
linked to the constituent mass of quarks introduced
in the context of  naive quark model \cite{KOK}. It is this type
of behavior of dynamical mass which is observed on the
lattice \cite{Aoki}.\\

As soon as the chiral symmetry is dynamically broken at
low momenta, then necessarily  appear Goldstone bosons
which couple to constituent quarks \cite{GEORGI}. This
property is most transparently illustrated by the
sigma-model \cite{Levy} and Nambu and Jona-Lasinio (NJL)
model \cite{Nambu}. The latter one suggests 
an insight into chiral symmetry breaking, constituent
mass generation, Nambu-Goldstone bosons as collective 
quark-antiquark modes. This model is
known to be very  successfull in the
low-lying meson spectrum, for reviews see \cite{reviews}.
It approximates a smooth drop of the dynamical
mass $M(p)$ by a step function, see Fig. 1
(because of a local character of the effective 4-fermion
interaction in this model). Thus at momenta
below chiral symmetry breaking scale $\Lambda_\chi$, which
corresponds to the ultraviolet cut-off  within the Nambu and Jona-Lasinio
model, the adequate effective degrees of freedom are the
constituent quarks and Goldstone bosons coupled to
each other.\\

It is instructive to review some of the basic properties
of the NJL model, which is nothing else but Bardeen-Cooper-Scrieffer
theory of
superconductivity, extended to explain chiral symmetry
breaking and dynamical mass generation.\\

Any scalar gluonic interaction between current quarks, which
is responsible for the chiral symmetry breaking in QCD, in the
{\it local} approximation is given by the 4-fermion operator
$(\bar \psi \psi)^2$. Because of the underlying chiral invariance
in QCD this interaction should be necessarily accompanied by
the interaction
 $(\bar \psi i\gamma_5 \vec \tau \psi)^2$ with the same strength.
Thus any generic Hamiltonian density in the local approximation is
given by (contains as a part) the NJL interaction model 
(for simplicity we restrict discussion 
 to the u,d flavour sector and to the chiral limit):

\begin{equation}
H= -G\left[(\bar \psi \psi)^2 + 
(\bar \psi i\gamma_5 \vec \tau \psi)^2\right]
\label{NJL}.
\end{equation}

\noindent
If the strength of the interaction $G$ exceeds some
critical level, then the nonlinear {\it gap} equation

\begin{equation}
M = -2G<\bar \psi \psi> 
\label{gap1},
\end{equation}

\begin{equation}
<\bar \psi \psi> = - \frac{N_c}{\pi^2}
\int_0^{\Lambda_\chi} d|{\bf p}|  {\bf p}^2 \frac{M}{\sqrt({\bf p}^2 + M^2)} 
\label{gap2}
\end{equation}

\noindent
admits a nontrivial solution $ <\bar \psi \psi> \not= 0$, which means
that the initial vacuum becomes rearranged and instead of the
Wigner-Weyl mode of chiral symmetry one obtains the Nambu-Goldstone
one. Thus the appearance of the quark condensate is equivalent
to the appearance of the gap in the spectrum of elementary excitations
in QCD (i.e. of the constituent mass $M$). Hence the constituent quark
is  a quasiparticle in the Bogoliubov sense,
i.e. is a coherent superposition of the bare particle-hole excitations.
 In terms
of the noninterating constituent quarks the vacuum is again trivial,
but contains a gap 2M. The treatment of the scalar interaction 
between bare quarks in the vacuum in
the Hartree-Fock (mean field) approximation, from which the gap
equation (\ref{gap1})-(\ref{gap2}) is obtained, is equivalent to the
vacuum of {\it noninteracting} constituent quarks. In the vacuum state
the second term of (\ref{NJL}) does
not contribute to the constituent quark self-energy in
the mean field approximation. Contributions beyond the mean field
approximation (i.e. constituent quark self energy due to pion
and sigma loops, which are higher order effects in the $1/N_c$ expansion,
see, e.g., \cite{D})
do not violate significantly the qualitative picture of
the vacuum in the mean field approximation.\\

In the systems that contain valence quarks on the top
of the vacuum (hadrons) the situation is qualitatively
different. In this case the valence quarks interact not only
with the vacuum condensates (which provides their constituent mass),
but also to each other. The latter interaction strongly depends
on the quantum numbers of hadrons.
 In the pseudoscalar-isovector quark-antiquark
system the second term of (\ref{NJL}), when iterated, exactly
compensates the $2M$ energy, supplied by the first term in the vacuum, 
and consequently
there appear massless pions as Nambu-Goldstone bosons.
This means that while for all other hadrons the chiral symmetry
breaking implies a presence of the gap in the excitation energy
($3M$ for baryons and $2M$ for mesons), this is not the case
for pions which represent a gapless excitation over QCD vacuum.
 Iteration
of the first term in the scalar-isoscalar quark-antiquark system
produces a weakly bound $\sigma$-meson with the mass $2M$. In the 
quark-quark systems, i.e. in baryons, iteration of the first
and second terms in the  $qq$ t-channel leads to the $\pi,\sigma$
exchange interactions between valence constituent quarks, which
reprtesents effects of the strong vacuum polarization \cite{GV}. The
pion-like exchange interactions between valence quarks in baryons
provide the $N-\Delta$ splitting and other low-lying baryon excitations
\cite{GR,GPVW}. So all interactions between valence quarks 
(i.e. beyond the
mean field approximation) can be reffered to as "residual", though
they are very strong. In baryons to these belong $\pi,\sigma$
exchanges and effective confining interaction between valence
quarks.  Which is the role of these residual interactions for the chiral symmetry restoration (breaking)
phase transition?\\

Consider first the chiral symmetry restoration in the vacuum
at high temperatures. At zero temperature the vacuum condensate
is given by (\ref{gap2}). At finite temperature it is affected
and given by

\begin{equation}
<\bar \psi \psi> = - \frac{N_c}{\pi^2}
\int_0^{\Lambda_\chi} d|{\bf p}|  {\bf p}^2 \frac{M}{E_p} [1-2n(p)] 
\label{gap3},
\end{equation}

\noindent
where $E_p =\sqrt({\bf p}^2 + M^2)$ and $n(p)$ is the 
Fermi-Dirac distribution function for  quarks and antiquarks

\begin{equation}
n(p) = \frac{1}{1 + e^{\frac{E_p}{T}}} 
\label{FD}.
\end{equation}

\noindent
In the latter equation $T$ is the temperature. 
At some critical temperature, $T_c$, the nontrivial gap solution
disappears and the chiral symmetry becomes restored. Physically
this is because the thermal excitations of quarks and antiquarks
lead to the Pauli blocking of the levels which are necessary
for the formation of the condensate. The formal (mathematical) reason
is that the quark distribution
function  $n(p)$ is pushed out from the $p=0$ point and becomes broad
and thus affects the gap equation so  that the self-consistent
(gap) solution disappears.
However, there could be other physical reason for the broadening
of the momentum
distribution. This physical reason is the strong
residual interactions between valence constituent quarks in baryons.\\

Consider the system that contains interacting valence quarks
on the top of the vacuum - baryons.
The residual interaction effect can be obtained from the
solution of the Schr\"odinger equation in the three-quark
systems \cite{GPVW}. As outcome one gets the quark
wave function (in the center of mass system) and this wave
function, being squared and properly normalized should
substitute $2n(p)$ in eq. (\ref{gap3}). 
For the ground state 3q system ($N$ and
$\Delta$) the one-quark wave function in momentum representation
is given by the narrow bell-shaped curve and does not
affect much the gap equation (\ref{gap3}). However, the
higher the radial  excitation of the baryon is, the larger is
the average kinetic energy
(momentum) of constituent quarks in baryons , i.e. the momentum distribution becomes broader
and broader. This momentum distribution is pushed out from the
inner part by the nodes of the wave function.
This means that at some baryon excitation energy the nontrivial
gap solution of eqs. (\ref{gap3})-(\ref{gap1}) should vanish
and the chiral symmetry should be restored. While numerically
this effect should be similar to chiral symmetry restoration
 at high temperature\footnote{In both cases the critical quantity is
average kinetic energy of quarks.}, physically {\it it is completely
different as provided by the residual interactions but not by
the thermal distributions.} \\

In the following I shall use for a qualitative estimate the 
fact that the chiral symmetry breaking scale is in the
region 

\begin{equation}
m_\rho < \Lambda_\chi < 4\pi f_\pi.
\label{scale}
\end{equation}

\noindent
For constituent mass I shall take  $M \sim 340$ MeV,
the value which is known from 60th and which is also obtained
in the recent lattice measurements \cite{Aoki,Hess}.
The  root mean square momentum of constituent quarks
for all low-lying baryons in $N$ and $\Delta$ spectra
(with masses $\leq 1.7 - 1.8$ GeV)
can be extracted from the wave
functions obtained from the fit to masses  in
dynamical semirelativistic calculation \cite{GPVW}.
This momentum falls into the range 500-700 MeV, which is
below $\Lambda_\chi$. This a-posteriori justifies a language
of constituent quarks that interact via GBE  and
are subject to confinement in the low-lying baryons. 
However, this momentum is not
very much below $\Lambda_\chi$. Hence,  baryons with mass
of 2 GeV and higher where the average momentum of quarks is larger,
should be in the region of the phase transition and 
a share of the Wigner-Weyl mode will
be big enough to ensure the appearance of the near parity
doublets.\\

This estimate is very consistent with the temperature of the
phase transition , which is known from
the Monte-Carlo lattice calculations, $T_c \sim 150$ MeV 
\cite{CT}, or the most
recent $T_c \sim 170 - 190$ MeV \cite{Karsch}. Above this
critical temperature, i.e. in the chiral restored phase,
the average kinetic energy of current quarks is above
$3T_c$ \footnote{This estimate is obtained
with the $\sim e^{-p/T}$ distribution function for current
quarks. With the Fermi-Dirac distribution function it is slightly
above $3T_c$.}. In all low-lying light baryons with mass below
1.7 - 1.8 GeV the
average kinetic energy of constituent quarks is in the
region  260 - 480 MeV, which is just below the critical one.
 Basing on these
simple arguments it is tempting to assume that one
observes a phase transition in the upper part of the light
baryon spectrum. If chiral and deconfinement phase
transitions coincide, the conclusion should be that the
highly excited baryons with masses above some critical value
(where the phase transition is completed) should not exist
because deconfinement phase transition should be dual to a
very extensive string breaking at big separations of colour
sources (colour screening). 
Whether this point corresponds to approximately 2.5 GeV or
higher should be answered by future experiments on high
baryon excitations. The phase transition can be rather
broad because of the explicit chiral symmetry breaking
by the nonzero value of current quark masses.\\

There is a couple of the well confirmed  states
 $N(2600), \frac{11}{2}^-$ and 
$\Delta(2420), \frac{11}{2}^-$, in which case the parity partners 
are absent \cite{Caso}. Thus it will be rather important to try to find
them experimentally.\\

A few comments about the parity doubling within the potential
models that attempt to describe the highly lying baryons
are in order. The models that rely on  confinement potential
cannot explain an appearance of the systematic parity doublets.
This is apparent for the harmonic confinement. The parity of
the state is determined by the number $N$  of the harmonic
excitation quanta in the 3q state. The ground states (N=0) are
of positive parity, all baryons from the $N=1$ band are of negative
parity, baryons from the $N=2$ band have a positive parity irrespective
of their angular momentum, etc. However, the number of states in
the given band rapidly increases with $N$. This means that such a
model cannot provide an equal amount of positive and negative
parity states, which is necessary for parity doubling, 
irrespective of other residual interactions between
quarks in such a model. Similar problem persists with the
linear confinement in 3q system.\\ 

While all vacancies from the $N=0$ and $N=1$ bands are filled
in in nature, such a model, extrapolated to the N=3 and higher bands 
predicts
a very big amount of states, which are not observed (the so called
missing resonance problem). According to the explanation suggested
in the present paper the chiral restoration phase transition takes
place at excitation energies typical for the $N=3$ band (and somewhat to
highest states from the $N=2$ band). If correct, it would mean that
description of  baryons in this transition region 
 in terms of constituent quarks becomes inappropriate.\\

The model that rely on the pure color Coulomb interaction between
quarks also cannot provide the systematical parity doubling. While
it gives an equal amount of the positive and negative parity
single quark states in the $n=2,4,...$ bands (e.g. $2s-2p$, or
$4s-4p,~ 4d-4f$), the number of the positive parity states
is always bigger in the $n=1,3,5,...$ bands.\\

Thus it is very important experimental task to verify whether
the "missing" resonances exist or not and whether the
upper part of the light baryon spectrum exhibits {\it systematical}
parity doublet patterns.\\

What about meson spectra? Here the strong residual interaction
due to Goldstone boson exchange is absent (it is impossible
in the quark-antiquark pairs), i.e. description of excited
mesons in terms of the constituent quarks interacting via chiral fields
is not possible and thus the arguments above
cannot be applied here. This is perhaps a reason for why there 
are no parity doubling patterns in meson spectra. If correct,
it then means that the baryon spectrum suggests a unique
opportunity to observe the chiral restoration phase transition.\\

I am indebted to the nuclear theory groups of KEK-Tanashi
and Tokyo Institute of Technology for a warm hospitality.
This work is supported by a foreign guestprofessorship
program of the Ministry of Education, Science, Sports and
Culture of Japan.

{\bf Figure captions}

Fig.1  A schematic behaviour of the dynamical mass of 
quarks as a function of their momenta as it is expected
in reality (solid line) and as it is approximated in the Nambu and 
Jona-Lasinio  as well as in chiral quark models (dashed line).

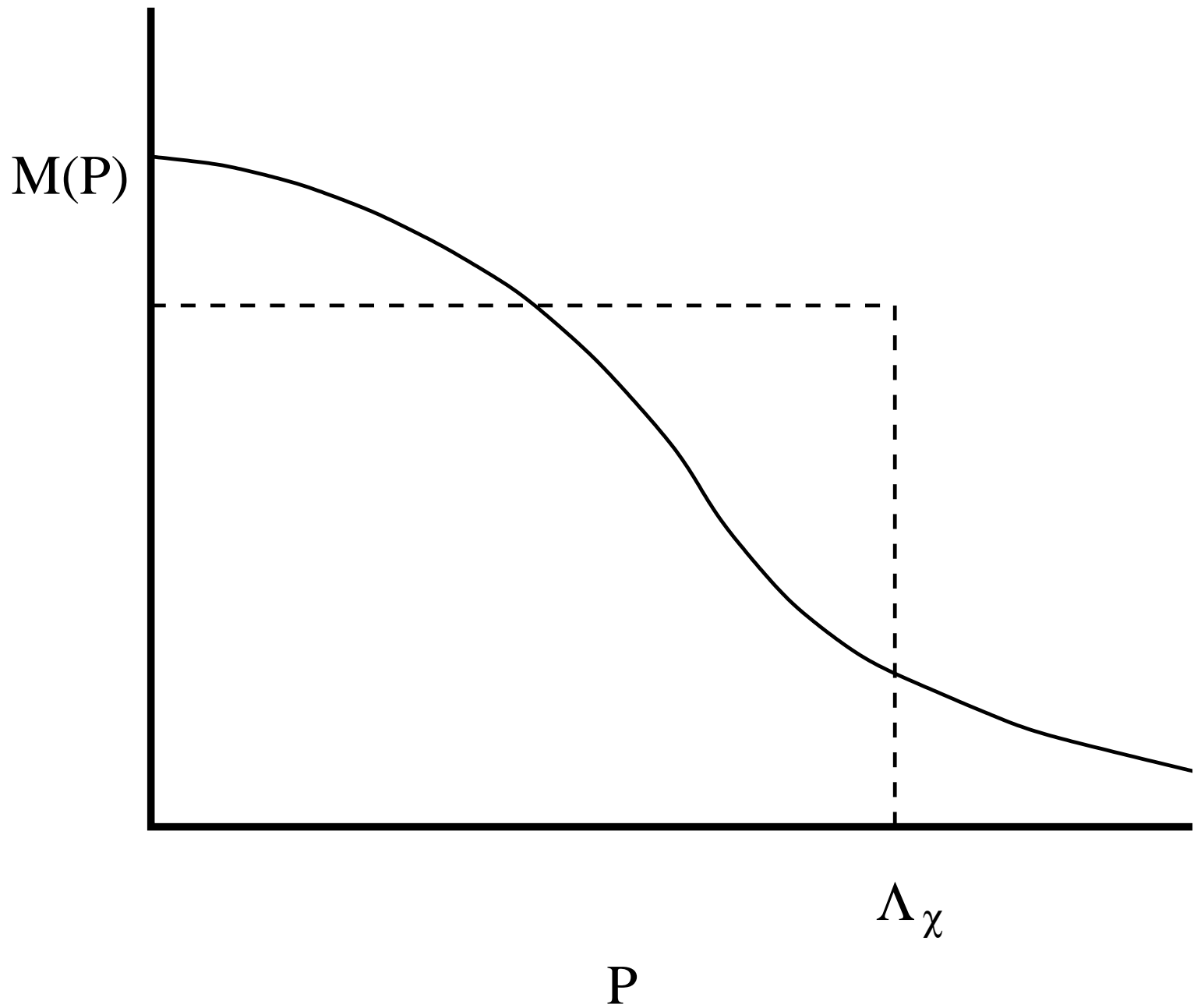
\end{document}